\documentclass[11pt,a4paper,reqno]{smfart}
\usepackage{amsthm,amsmath,amsfonts,amssymb,amsxtra,appendix,bookmark,dsfont,latexsym,bm,hyperref,delarray,color,euscript,amsgen,amsbsy,amsopn,amscd,latexsym,mathrsfs,smfhyperref}

\makeatletter

\usepackage{hyperref}

\makeatletter
\def\@tocline#1#2#3#4#5#6#7{\relax
  \ifnum #1>\c@tocdepth 
  \else
    \par \addpenalty\@secpenalty\addvspace{#2}%
    \begingroup \hyphenpenalty\@M
    \@ifempty{#4}{%
      \@tempdima\csname r@tocindent\number#1\endcsname\relax
    }{%
      \@tempdima#4\relax
    }%
    \parindent\z@ \leftskip#3\relax \advance\leftskip\@tempdima\relax
    \rightskip\@pnumwidth plus4em \parfillskip-\@pnumwidth
    #5\leavevmode\hskip-\@tempdima
      \ifcase #1
       \or\or \hskip 1em \or \hskip 2em \else \hskip 3em \fi%
      #6\nobreak\relax
      \dotfill
      \hbox to\@pnumwidth{\@tocpagenum{#7}}
    \par
    \nobreak
    \endgroup
  \fi}
\makeatother

\newtheorem{theorem}{Th\'eor\`eme}[section]
\newtheorem{Enon}[theorem]{Enonc\'e}
\newtheorem{lemma}[theorem]{Lemme}

\theoremstyle{definition}

\newtheorem{example}[theorem]{Exemple}
\newtheorem{remark}[theorem]{Remarque}

\newcommand{\ii}{\infty}
\newcommand{\im}{\mathrm{i}}
\newcommand\R{{\ensuremath {\mathbb R} }}
\newcommand\C{{\ensuremath {\mathbb C} }}
\newcommand\N{{\ensuremath {\mathbb N} }}

\newcommand\1{{\ensuremath {\mathds 1} }}

\renewcommand\phi{\varphi}

\newcommand{\bH}{\mathbb{H}}

\newcommand{\gH}{\mathfrak{H}}
\newcommand{\gF}{\mathfrak{F}}

\newcommand{\cE}{\mathcal{E}}

\newcommand{\cF}{\mathcal{F}}
\newcommand{\cN}{\mathcal{N}}
\newcommand{\cD}{\mathcal{D}}

\newcommand{\cZ}{\mathcal{Z}}

\renewcommand{\epsilon}{\varepsilon}

\newcommand{\norm}[1]{ \left| \! \left| #1 \right| \! \right| }
\DeclareMathOperator{\tr}{{\rm Tr}}
\DeclareMathOperator{\Tr}{{\rm Tr}}

\renewcommand{\ge}{\geqslant}
\renewcommand{\le}{\leqslant}
\renewcommand{\geq}{\geqslant}
\renewcommand{\leq}{\leqslant}
\renewcommand{\hat}{\widehat}
\renewcommand{\tilde}{\widetilde}

\newcommand{\dG}{\mathrm{d}\Gamma}

\newcommand{\rhoO}{\varrho_0}

\numberwithin{equation}{section}

\begin{document}

\title{Limites de champ moyen bosoniques \`a temp\'erature positive}



\author[N. Rougerie]{Nicolas Rougerie}
\address{Universit\'e Grenoble-Alpes \& CNRS,  LPMMC (UMR 5493), B.P. 166, F-38042 Grenoble, France}
\email{nicolas.rougerie@lpmmc.cnrs.fr}

\date{D\'ecembre 2018}

\renewcommand{\abstractname}{R\'esum\'e}

\begin{abstract}
Je pr\'esente des r\'esultats r\'ecents de m\'ecanique statistique quantique, obtenus avec Mathieu Lewin et Phan Th\`anh Nam. Nous consid\'erons une certaine limite de champ moyen pour l'ensemble grand-canonique d'un gaz de bosons \`a temp\'erature positive. Dans cette limite, les matrices de densit\'e r\'eduites de la th\'eorie quantique convergent vers leurs pendants en th\'eorie des champs classiques, donn\'ees par une mesure de Gibbs non-lin\'eaire. Nous traitons en particulier de cas o\`u cette derni\`ere doit \^etre d\'efinie via une proc\'edure de renormalisation. La renormalisation correspondante au niveau du probl\`eme grand-canonique de d\'epart, avec des champs non commutatifs, est une des difficult\'es majeures.
\end{abstract}

\maketitle

\tableofcontents


\section{Introduction}

Le but de cette introduction est de d\'efinir de mani\`ere informelle les mots du titre. La lectrice qui les conna\^it d\'ej\`a tous pourra sauter directement \`a la section suivante.

\medskip

\noindent\textbf{Qu'est-ce qu'une \emph{limite} de champ moyen ?} Par convention souvent tacite il s'agit d'un r\'egime asymptotique o\`u l'on peut justifier une \emph{approximation} de champ moyen. C'est-\`a-dire, \'etant donn\'e un syst\`eme physique fait de nombreuses particules interagissantes via un certain Hamiltonien microscopique, d\'emontrer que les propri\'et\'es macroscopiques du syst\`eme peuvent s'approcher via une approximation en apparence assez crue: supposer que toutes les particules sont ind\'ependantes et ind\'ependamment distribu\'ees, et calculer la distribution typique d'une particule g\'en\'erique sous l'effet d'un champ auto-consistant (d\'ependant de la distribution g\'en\'erique elle-m\^eme) qui est une repr\'esentation ``en moyenne'' du champ ressenti par une particule g\'en\'erique \`a cause de l'influence de toutes les autres. 

En m\'ecanique statistique classique, l'\'etat d'un ensemble de $N$ particules dans l'espace $\R^d$ est d\'ecrit par une mesure de probabilit\'e $\mu (x_1,\ldots,x_N;p_1,\ldots,p_N)$ donnant la densit\'e de probabilit\'e pour que les particules occupent les positions $x_1,\ldots, x_N \in \R^d$ et aient les vitesses/moments/impulsions\footnote{Je serais dans ce texte tr\`es vague quant aux unit\'es et dimensions.} $p_1,\ldots,p_N \in \R^d$. Si les particules sont toutes du m\^eme type elles sont par d\'efinition indiscernables, et donc $\mu$ doit \^etre invariante sous tous les \'echanges de jeux de coordonn\'ees $(x_i,p_i) \leftrightarrow (x_j,p_j)$. La loi la plus simple que l'on puisse alors imaginer  correspond \`a des particules ind\'ependantes et identiquement distribu\'ees:
\begin{equation}\label{eq:MF classique}
\mu (x_1,\ldots,x_N;p_1,\ldots,p_N) = \prod_{j=1} ^N f(x_j;p_j).  
\end{equation}
Justifier une approximation de champ moyen consiste \`a prouver que l'on peut l\'egitimement faire cette simplification pour calculer au moins les grandeurs macroscopiques du syst\`eme. L'importance de l'enjeu est \'evident si l'on se rappelle qu'essentiellement \emph{toute} la physique aux \'echelles macroscopiques (th\'eorie cin\'etique des gaz, m\'ecanique des fluides ...) est bas\'ee sur ce genre d'approximation au niveau microscopique.

\medskip

\noindent\textbf{Qu'est-ce qu'une limite de champ moyen \emph{bosonique} ?} Tout d'abord le terme ``bosonique'' ne fait sens que dans un cadre quantique. En m\'ecanique statistique quantique, l'\'etat d'un ensemble de $N$ particules dans l'espace $\R^d$ est d\'ecrit par un op\'erateur \`a trace positif $\Gamma_N$ agissant sur l'espace de Hilbert
\begin{equation}\label{eq:Hilbert sans sym}
 \gH_N := L^2 (\R^{dN},\C)= \bigotimes^N L^2 (\R^d,\C). 
\end{equation}
On normalise $\Gamma_N$ pour que sa trace soit $1$. L'interpr\'etation de ce formalisme est un peu moins ais\'e: le principe d'incertitude d'Heisenberg postule que les observables de position et de moment ne commutent pas entre elles. En fait, on postule que $x_j,p_j$ ne sont plus des (multiplications par des) nombres mais des op\'erateurs ayant un commutateur
\begin{equation}\label{eq:Heisenberg}
 [x_j,p_j] = \mathrm{i}\hbar  
\end{equation}
avec $\hbar$ la constante de Planck r\'eduite. Discuter en d\'etail de ce postulat nous emm\`enerait trop loin. Contentons-nous d'observer que si l'on d\'ecide que $x_j$ est l'op\'erateur de multiplication par la coordonn\'ee $x_j$ sur l'espace~\eqref{eq:Hilbert sans sym} et $p_j$ l'op\'erateur de multiplication en Fourier par $p_j:= \hbar k_j$ avec $k_j$ la variable duale de $x_j$, nous construisons effectivement une repr\'esentation\footnote{De mani\`ere \'equivalente, on peut voir le moment comme $p_j=-\im \hbar \nabla_{x_j}$.} des relations de commutation~\eqref{eq:Heisenberg}.    

L'interpr\'etation de notre op\'erateur \`a trace $\Gamma_N$ passe alors par ses noyaux int\'egraux $\Gamma_N(x_1,\ldots,x_N;y_1,\ldots,y_N)$ et $\widehat{\Gamma_N} (p_1,\ldots,p_N;q_1,\ldots,q_N)$ en variables directes et de Fourier respectivement:
$$ \mu_N ^{\rm pos}(x_1,\ldots,x_N) := \Gamma_N(x_1,\ldots,x_N;x_1,\ldots,x_N)$$
repr\'esente la densit\'e de probabilit\'e des particules dans l'espace des positions, et 
$$\mu_N ^{\rm mom} (p_1,\ldots,p_N) := \widehat{\Gamma_N} (p_1,\ldots,p_N;p_1,\ldots,p_N)$$
la densit\'e de probabilit\'e des particules dans l'espace des moments. 

Pour des particules indiscernables, il nous faut bien s\^ur imposer que ces distributions soient invariantes par permutation des coordonn\'ees des particules. Incoporer cette contrainte de mani\`ere logiquement consistante est une longue t\^ache\footnote{Et lorsque l'on se d\'ecide \`a la mener r\'eellement \`a bien, on d\'ecouvre des surprises fort int\'eressantes~\cite{LeiMyr-77}.} dont nous donnerons seulement la conclusion: il y a deux choix physiquement pertinents, donnant une dichotomie fondamentale de la nature (\`a l'origine, entre autres, de la table de Mendele\"iev et du principe du laser). 

Soit les particules d\'ecrites par $\Gamma_N$ sont des bosons et on impose alors que $\Gamma_N$ agisse seulement sur la version sym\'etris\'ee de~\eqref{eq:Hilbert sans sym}:
\begin{equation}\label{eq:Hilbert sym}
 \gH_N := L^2_{\rm sym} (\R^{dN},\C) = \bigotimes_{\rm sym} L^2 (\R^d,\C)
\end{equation}
c'est-\`a-dire qu'il vit uniquement sur le sous-espace des fonctions sym\'etriques satisfaisant 
$$ \Psi (x_{\sigma(1)},\ldots,x_{\sigma(N)}) = \Psi (x_1,\ldots,x_N)$$
pour toute permutation $\sigma$. 

Soit les particules d\'ecrites par $\Gamma_N$ sont des fermions et on impose alors que $\Gamma_N$ agisse seulement sur la version anti-sym\'etris\'ee de~\eqref{eq:Hilbert sans sym}:
\begin{equation}\label{eq:Hilbert asym}
 \gH_N := L^2_{\rm asym} (\R^{dN},\C) = \bigotimes_{\rm asym} L^2 (\R^d,\C)
\end{equation}
c'est-\`a-dire qu'il vit uniquement sur le sous-espace des fonctions anti-sym\'etriques satisfaisant 
$$ \Psi (x_{\sigma(1)},\ldots,x_{\sigma(N)}) = (-1) ^{\mathrm{sign} (\sigma)} \Psi (x_1,\ldots,x_N)$$
pour toute permutation $\sigma$, avec $\mathrm{sign} (\sigma)$ la signature de $\sigma$.  

On remarquera que pour des particules bosoniques, un ansatz ``particules ind\'ependantes'' inspir\'e de~\eqref{eq:MF classique} est l\'egitime, alors que les fermions ont toujours des corr\'elations minimales (principe d'exclusion de Pauli) qui les emp\^eche d'\^etre ind\'ependantes. Une \emph{limite de champ moyen bosonique} est donc une limite o\`u l'on pourra justifier de la validit\'e d'un ansatz de forme 
\begin{align}
\Gamma_N &= \gamma ^{\otimes N} \nonumber\\
\Gamma_N (x_1,\ldots,x_N;y_1,\ldots,y_N) &= \prod_{j=1} ^N \gamma (x_j;y_j)\label{eq:MF bosons}
\end{align}
pour d\'ecrire le syst\`eme \'etudi\'e, au premier ordre dans la limite $N\to \infty$. Ici $\gamma$ est un op\'erateur sur $L^2 (\R^d)$ et la deuxi\`eme \'equation est dans le sens des noyaux int\'egraux. Un ansatz de cette forme a la sym\'etrie bosonique si et seulement si $\gamma = |u\rangle \langle u| = u (x) \overline{u (y)}$ est un projecteur orthogonal (\'etat pur) sur une fonction $u$ normalis\'ee dans $L^2$. 

\medskip

\noindent\textbf{Qu'est-ce qu'une limite de champ moyen bosonique \emph{\`a temp\'erature positive} ?} Commen\c{c}ons par rappeler que les temp\'eratures sont par d\'efinition positives ou nulles, et par expliquer le cas de la temp\'erature nulle. Dans cette situation, les \'etats d'\'equilibre d'un syst\`eme sont donn\'es par la minimisation de l'\'energie (associ\'ee \`a un op\'erateur Hamiltonien) correspondante. On sait montrer~\cite{Lewin-ICMP,Lewin-XEDP-12,LieSeiSolYng-05,Rougerie-spartacus,Rougerie-LMU} avec une g\'en\'eralit\'e satisfaisante que l'ansatz~\eqref{eq:MF bosons} est valide pour de tels minimiseurs (en un sens appropri\'e qu'il faut discuter assez soigneusement). On sait aussi montrer~\cite{BenPorSch-15,Golse-13,Schlein-08} que la factorisation~\eqref{eq:MF bosons} est (\`a nouveau dans un sens appropri\'e) pr\'eserv\'ee par l'\'evolution temporelle ad\'equate (flot de Schr\"odinger \`a $N$ corps associ\'e au Hamiltonien). 

Alors, quid des \emph{limite de champ moyen bosoniques \`a temp\'erature positive} ? Dans ce texte, je d\'esignerais par ce terme une limite o\`u on doit aller au del\`a de l'ansatz~\eqref{eq:MF bosons} pour d\'eterminer les \'etats d'\'equilibre d'un syst\`eme. A temp\'erature non nulle, il convient de minimiser l'\'energie libre: \'energie $-$ temp\'erature $\times$ entropie. On sait beaucoup moins de choses avec un degr\'e de rigueur math\'ematique suffisant dans ce cas (voir cependant~\cite{BetUel-10,Seiringer-06,Seiringer-08,SeiUel-09,Yin-10,DeuSeiYng-18}). Le but de cette note est de discuter une situation o\`u l'on obtient en fait une superposition statistique (convexe) non-triviale d'\'etats factoris\'es de forme~\eqref{eq:MF bosons}. L'entropie favorise cet ``\'etalement'' dans l'espace des configurations possibles.   

\medskip

\noindent\textbf{Annonce du th\`eme.} Il s'agit donc de pr\'esenter une variation sur le th\`eme des limites de champ moyen bosoniques, au sens o\`u l'on ira au-del\`a de l'ansatz iid habituel. Le cas des temp\'eratures nulles, o\`u l'ansatz iid est suffisant, est beaucoup mieux connu, et je renvoie aux r\'ef\'erences ci-dessus. Les questions que l'on aura \`a se poser ici seront les suivantes: 

\smallskip

\noindent $\bullet$ Quelle est la nature de la superposition statistique d'op\'erateurs de forme~\eqref{eq:MF bosons} appropri\'ee \`a notre contexte ? 
 
\smallskip

\noindent $\bullet$ Comment la d\'efinir pr\'ecis\'ement ? On verra que le choix naturel am\`ene \`a des difficult\'es d'analyse: des probl\`emes de renormalisation se posent, rappelant celles apparaissant en th\'eorie constructive des champs~\cite{DerGer-13,GliJaf-87,Simon-74,Summers-12}, dans la th\'eorie de Cauchy probabiliste pour l'\'equation de Schr\"odinger non-lin\'eaire~\cite{LebRosSpe-88,Bourgain-96,Bourgain-97,BurThoTzv-09,BurThoTzv-10,CacSuz-14,OhTho-15}, dans l'\'etude de l'\'equation de la chaleur non-lin\'eaire stochastique ~\cite{BouDebFuk-17,PraDeb-03,Hairer-14,Kupiainen-16,MouWeb-15,RocZhuZhu-16,TsaWeb-16}.
 
\smallskip

\noindent $\bullet$ Comment effectuer la limite de champ moyen liant la superposition statistique aux \'etats du syst\`eme physique originel ?
 
\smallskip

\noindent $\bullet$ Comment contr\^oler la renormalisation lorsqu'elle est n\'ecessaire \`a la d\'efinition de l'objet limite ? Des termes de renormalisation sont alors n\'ecessaires au niveau du probl\`eme \`a $N$ corps, et il faut pouvoir passer \`a la limite dans ceux-ci \'egalement.


\section{Limites de champ moyen \`a temp\'erature nulle}

Les \'enonc\'es de cette section sont volontairement gard\'es vagues. On renvoie \`a la lit\'erature cit\'ee pour leur mise en forme pr\'ecise.

\subsection{Mod\`ele de d\'epart: Schr\"odinger $N$-corps} 

Le Hamiltonien pour $N$ particules quantiques non relativistes, interagissant via un potentiel de paire $w: \R^d \mapsto \R$ et avec un potentiel ext\'erieur $V:\R^d \mapsto \R$ s'\'ecrit\footnote{On pourrait \'eventuellement inclure un champ magn\'etique et des effets pseudo-relativistes, mais restons aussi simples que faire se peut.} (en choisissant les unit\'es telles que $m=1/2, \hbar = 1$): 
\begin{equation}\label{eq:N body Schro}
H_{N,\lambda} = \sum_{j=1} ^N \left(-\Delta_{x_j} + V (x_j) \right) + \lambda \sum_{1\leq i < j \leq N} w (x_i - x_j) 
\end{equation}
avec $\lambda \in \R$ une constante de couplage. On suppose en g\'en\'eral que $w(x) = w(-x)$, voir que $w(x) = w (|x|)$ est une fonction radiale. Dans ce texte on se focalisera plus particuli\`erement sur le cas d'un potentiel ext\'erieur confinant: 
$$ V(x) \underset{x\to \infty} \longrightarrow + \infty.$$
Comme discut\'e ci-dessus, pour des particules bosoniques il convient de ne consid\'erer l'action de cet op\'erateur que sur l'espace \`a $N$ corps sym\'etrique~\eqref{eq:Hilbert sym}. Sous des hypoth\`eses standard, $H_N$ est auto-adjoint avec le m\^eme domaine que la version sans interaction ($\lambda=0$). Les objets d'int\'er\^et sont le spectre et les fonctions propres associ\'ees, trouv\'ees comme valeurs min-max de l'\'energie
\begin{equation}\label{eq:N corps ener}
 \cE_{N,\lambda} [\Psi_N] = \langle \Psi_N |H_{N,\lambda}| \Psi_N  \rangle_{L^2 (\R^{dN})}
\end{equation}
pour $\Psi_N \in L^2_{\rm sym} (\R^{dN})$ sous la contrainte de masse 
\begin{equation}\label{eq:N corps masse}
\int_{\R^{dN}} |\Psi_N| ^2 = 1.  
\end{equation}
L'\'evolution en temps d'une fonction d'onde initiale $\Psi_N^0$ est donn\'ee par le flot de Schr\"odinger $N$ corps
\begin{equation}\label{eq:N corps Schro}
\begin{cases}
 \im \partial_t \Psi_N = H_{N,\lambda} \Psi_N \\
 \Psi_N (t=0) = \Psi_N^0.
\end{cases} 
\end{equation}
En quoi tout ceci est-il sp\'ecifique \`a la temp\'erature nulle ? Essentiellement parce que tous les \'etats $\Gamma_N$ au sens de la section pr\'ec\'edente sont ici des \'etats purs, i.e. des projecteurs orthogonaux:
$$ \Gamma_N = |\Psi_N \rangle \langle \Psi_N|$$
ayant une entropie nulle (l'entropie quantique d'un \'etat est toujours positive ou nulle.)

\subsection{Limite de champ moyen} 

Deux probl\`emes sont plus particuli\`erement int\'eressants: 

\smallskip

\noindent$\bullet$ minimisation de l'\'energie~\eqref{eq:N corps ener} par rapport \`a $\Psi_N$ sous la contrainte de masse~\eqref{eq:N corps masse}. A temp\'erature nulle, les minimiseurs (ou plus pr\'ecis\'ement les projecteurs orthogonaux associ\'es) sont les \'etats d'\'equilibre du syst\`eme. 

\smallskip

\noindent$\bullet$ l'\'evolution suivant un flot de type~\eqref{eq:N corps Schro} de minimiseurs d'\'energie. Typiquement (et cela correspond \`a une large classe d'exp\'eriences) on prend comme donn\'ee initiale $\Psi_N^0$ un minimiseur de l'\'energie correspondant \`a $H_{N,\lambda}$ pour un certain $V$, et on le fait \'evoluer suivant le flot~\eqref{eq:N corps Schro} correspondant \`a un autre $V$.

\smallskip

On a alors les \'enonc\'es g\'en\'eriques suivants, que je ne chercherai pas \`a formuler rigoureusement: 

\begin{Enon}[\textbf{Fondamental \`a temp\'erature nulle}]\label{eno:fond}\mbox{}\\
Sous des hypoth\`eses raisonnables sur $V$ et $w$, soit $E (N,\lambda)$ 
le minimum de l'\'energie~\eqref{eq:N corps ener} sous la contrainte de masse~\eqref{eq:N corps masse}. On a, dans la limite 
\begin{equation}\label{eq:MF lim 0}
 N \to \infty, \qquad \lambda = \frac{g}{N} 
\end{equation}
avec $g$ une constante fixe, le r\'esultat 
\begin{equation}\label{eq:MF ener lim}
E (N,\lambda) = \min\left\{ \cE_{N,\lambda} [u ^{\otimes N}] \: \big| \: \int_{\R^d} |u| ^2 = 1 \right\} (1 + o (1)).
\end{equation}
Si de plus le minimum dans le membre de droite est atteint par un unique $u\in L^2 (\R^d)$, alors, tout minimiseur $\Psi_N$ de~\eqref{eq:N corps ener} satisfait 
$$ \Psi_N \approx u^{\otimes N}$$
en un certain sens dans la limite~\eqref{eq:MF lim 0}. 
\end{Enon}

Quelques commentaires:

\smallskip

\noindent(1) En termes un peu vagues, on peut restreindre la minimisation de l'\'energie aux fonctions factoris\'ees (ansatz de champ moyen, particules iid)
$$ u ^{\otimes N} (x_1,\ldots,x_N) = \prod_{j=1} ^N u(x_j).$$
Les minimiseurs des probl\`emes restreints et g\'en\'eraux coincident dans la limite $N\to \infty$.

\smallskip

\noindent(2) La fa\c{c}on de choisir $\lambda$ dans~\eqref{eq:MF lim 0} est motiv\'ee par des consid\'erations un peu ad-hoc ($N$ termes dans la premi\`ere somme de~\eqref{eq:N body Schro}, $\sim N^2$ dans la seconde, $\lambda$ choisi pour rendre les deux sommes comparables). Ce n'est pas la fa\c{c}on la plus physique d'obtenir une approximation de champ moyen, on pr\'ef\`erera souvent une limite dite dilu\'ee. Encore une fois, restons simples pour ce texte, et contentons-nous de pr\'eciser que ce cas (plus difficile !!) est \'egalement couvert par les techniques math\'ematiques existantes.

\smallskip

\noindent(3) Il y a une \textbf{tr\`es importante subtilit\'e conceptuelle} cach\'ee derri\`ere les mots ``en un certain sens'' \`a la fin de l'\'enonc\'e. Les fonctions $\Psi_N$ et $u^{\otimes N}$ ne sont PAS proches en norme $L^2$ par exemple. Je reporte \`a plus tard dans l'expos\'e la formulation rigoureuse de la limite de champ moyen pour des \'etats quantiques, qui n\'ecessite la notion de matrice de densit\'e r\'eduite.

\smallskip

Le choix de donn\'ees initales pour le probl\`eme dynamique est motiv\'e par l'\'enonc\'e pr\'ec\'edent: sous des hypoth\`eses g\'en\'eriques, un fondamental (minimiseur de l'\'energie) est factoris\'e. On souhaite voir si cette propri\'et\'e est pr\'eserv\'ee par l'\'evolution temporelle:

\begin{Enon}[\textbf{Dynamique \`a temp\'erature nulle}]\label{eno:dyn}\mbox{}\\
Sous des hypoth\`eses raisonnables sur $V$ et $w$, soit $\Psi_N (t)$ la solution de l'\'equation de Schr\"odinger \`a $N$ corps~\eqref{eq:N corps Schro}. Si en un sens appropri\'e
$$ \Psi_N (t=0) \approx u_0 ^{\otimes N}$$
pour un certain $u_0 \in L^2 (\R^d)$ alors, dans le m\^eme sens appropri\'e et dans la limite~\eqref{eq:MF lim 0}
\begin{equation}\label{eq:MF dyn lim}
 \Psi_N (t) \approx u(t) ^{\otimes N} 
\end{equation}
pour un certain $u(t)\in L^2 (\R^d)$.
\end{Enon}

Quelques commentaires:

\smallskip

\noindent(1) Je reste \`a nouveau vague sur le ``en un sens appropri\'e''. On peut toutefois mentionner que ce ``certain sens'' est le m\^eme que celui \`a la fin de l'\'enonc\'e~\ref{eno:fond}, de sorte que les deux \'enonc\'es se combinent \'el\'egamment. 
 
\smallskip

\noindent(2) L'approximation~\eqref{eq:MF dyn lim} est valable pour des temps $t$ fixes dans la limite~\eqref{eq:MF lim 0}. Obtenir des bornes efficaces sur la d\'ependance en $N$ du temps pour lequel l'approximation reste valable est un probl\`eme ouvert.

\subsection{Mod\`ele d'arriv\'ee: Schr\"odinger non-lin\'eaire} 

Les r\'esultats rappel\'es ci-dessus nous indiquent le mod\`ele de champ moyen appropri\'e: on l'obtient en rempla\c{c}ant les fonctions d'onde \`a $N$ particules $\Psi_N$ par des fonctions factoris\'ees $u^{\otimes N}$. On obtient alors la fonctionnelle d'\'energie de champ moyen (MF = mean-field) 
\begin{multline}\label{eq:MF func}
\cE^{\rm MF}_{g} [u] = \int_{\R^d} |\nabla u | ^2 + V |u| ^2 + \frac{g}{2} \iint_{\R^d \times \R^d} |u(x)|^2 w (x-y) |u(y)| ^2 dxdy \\ = N^{-1} \left\langle u^{\otimes N} | H_{N,\lambda} | u ^{\otimes N} \right\rangle  
\end{multline}
o\`u la deuxi\`eme \'egalit\'e demande le choix  
$$\lambda = \frac{g}{N-1}.$$
Le probl\`eme de minimisation de l'\'energie devient 
\begin{equation}\label{eq:MF ener}
E^{\rm MF}_g = \min\left\{ \cE_g ^{\rm MF}  [u] \: \big| \: \int_{\R^d} |u| ^2 = 1 \right\} 
\end{equation}
et il fournit l'\'energie et les minimiseurs associ\'es apparaissant dans l'Enonc\'e~\ref{eno:fond}.

Il est naturel que l'\'evolution correspondante soit donn\'ee par le flot Hamiltonien associ\'e \`a~\eqref{eq:MF ener}: 
\begin{align}\label{eq:MF dyn}
\im \partial_t u &= \partial_{\bar{u}} \cE^{MF}_g [u,\bar{u}] \nonumber\\
&= -\Delta u + V u + g (w\star |u|^2)u.
\end{align}
La fonction $u(t)$ apparaissant dans l'Enonc\'e~\ref{eno:dyn} est solution de cette \'equation de Schr\"odinger non-lin\'eaire,  que l'on peut \'egalement obtenir en supposant que la solution de~\eqref{eq:N corps Schro} reste factoris\'ee \`a temps positif (ce qui n'est pas vrai, mais au moins approximativement vrai selon l'Enonc\'e~\ref{eno:dyn}).

\section{Mod\`eles \`a temp\'erature positive}

\subsection{L'ensemble canonique et l'approximation de champ moyen}\label{sec:can}

A temp\'erature nulle, les \'etats d'\'equilibre (fondamentaux) minimisent l'\'energie. A temp\'erature positive, ils doivent trouver un compromis entre minimiser l'\'energie et maximiser l'entropie. On formalise ceci en introduisant la fonctionnelle d'\'energie libre (\'energie $-$ temp\'erature $\times$ entropie)
\begin{equation}\label{eq:N corps ener libre func}
\cF_{N,\lambda} [\Gamma_N] = \tr\left( H_{N,\lambda} \Gamma_N \right) + T \tr \left( \Gamma_N \log \Gamma_N \right).
\end{equation}
Ici $\Gamma_N$ est un \'etat (op\'erateur positif de trace $1$) sur l'espace sym\'etrique~\eqref{eq:Hilbert sym} et les traces sont prises sur cet espace. Noter que si $\Gamma_N = |\Psi_N \rangle \langle \Psi_N |$ est un \'etat pur, le premier terme devient~\eqref{eq:N corps ener} et le second vaut $0$. Le point est que pour une temp\'erature $T>0$, le second terme favorise un \'etat qui ne soit pas un projecteur. En fait, on peut r\'esoudre explicitement le probl\`eme de minimisation 
\begin{equation}\label{eq:N corps ener libre}
F_{N,\lambda} = \min \left\{ \cF_{N,\lambda} [\Gamma_N] \: | \: \Gamma_N \geq 0, \tr \Gamma_N = 1 \right\}. 
\end{equation}
On obtient comme minimiseur l'\'etat de Gibbs 
\begin{equation}\label{eq:N corps Gibbs}
\Gamma_{N,\lambda} = \frac{1}{\cZ_{N,\lambda}} \exp\left( -\frac{1}{T} H_N \right) 
\end{equation}
o\`u la fonction de partition $\cZ_{N,\lambda}$ normalise la trace et est reli\'ee \`a l'\'energie libre minimale par la relation 
$$ F_{N,\lambda} = - T \log \cZ_{N,\lambda}.$$

\medskip

Par quoi pourrait-on approximer~\eqref{eq:N corps Gibbs}, gardant une partie de l'esprit de la limite de champ moyen ? L'id\'ee est que~\eqref{eq:N corps Gibbs} doit \^etre vue comme une probabilit\'e sur les \'etats \`a $N$ particules $\Psi_N$:
$$ \left\langle \Psi_N | \Gamma_{N,\lambda} | \Psi_N\right\rangle = \frac{1}{\cZ_{N,\lambda}} \sum_{j} \exp\left(-\frac{E^j _N}{T}\right)\left|\left\langle \Psi_N^j | \Psi_N \right\rangle\right| ^2  
$$
o\`u les $E_N ^j,\Psi_N^j$ sont les valeurs propres, et fonctions propres associ\'ees, de $H_{N,\lambda}$ (restreint \`a l'espace sym\'etrique~\eqref{eq:Hilbert sym}). Modulo normalisation le poids 
$$\exp\left(-\frac{E^j _N}{T}\right)$$
est la probabilit\'e que le syst\`eme soit dans l'\'etat $\Psi_N^j$. 

Les r\'esultats de la section pr\'ec\'edente sugg\`erent que, dans une limite de champ moyen, seuls survivent les \'etats factoris\'es $\Psi_N = u ^{\otimes N}$. Il est donc naturel de chercher une mesure de probabilit\'es sur de telles fonctions, donc une mesure sur les fonctions \`a une particule $u\in L ^2 (\R^d)$. En invoquant \`a nouveau les r\'esultats \`a temp\'erature nulle, on aurait assez envie d'associer \`a la fonction $u^{\otimes N}$ le poids  
\begin{align*}
 \left\langle u ^{\otimes N} | \Gamma_{N,\lambda} | u ^{\otimes N} \right\rangle &= \frac{1}{\cZ_{N,\lambda}}   \left\langle u ^{\otimes N} | \exp\left( -\frac{1}{T} H_N \right)  | u ^{\otimes N} \right\rangle\\
 &\approx \frac{1}{\cZ_{N,\lambda}}  \exp\left( -\frac{1}{T} \left\langle u ^{\otimes N} | H_{N,\lambda} | u ^{\otimes N} \right\rangle\right)  \\
 &= \frac{1}{\cZ_{N,\lambda}}   \exp\left( -\frac{N}{T} \cE ^{\rm MF}_g [u] \right),  
\end{align*}
en faisant une manipulation assez formelle en deuxi\`eme \'etape: on esp\`ere que les $u^{\otimes N}$ qui seront charg\'es par la mesure sont quasiment des fonctions propres de $H_{N,\lambda}$. Vu que l'on voudrait obtenir une mesure de probabilit\'e, on aimerait ensuite d\'ecorer la derni\`ere ligne avec une mesure $du$ sur les fonctions $L^2 (\R^d)$. On arrive ainsi \`a un candidat ``naturel'' pour approximer~\eqref{eq:N corps Gibbs} dans la limite de champ moyen:
\begin{equation}\label{eq:NL Gibbs}
\boxed{\mu (du) := \frac{1}{z^{\rm MF}} \exp\left( -\frac{1}{t} \cE^{\rm MF}_g [u] \right) du}
\end{equation}
o\`u $t,g$ sont des param\`etres donnant une temp\'erature et une constante de couplage effectives respectivement, et $z^{\rm MF}$ une fonction de partition, normalisant l'objet pour en faire une mesure de probabilit\'e. 

Comme la lectrice peut s'en douter, il n'est pas si \'evident de donner un sens pr\'ecis \`a la formule~\eqref{eq:NL Gibbs}, en particulier \`a cause de la pr\'etendue ``mesure de Lebesgue sur $L^2 (\R^d)$'' $du$. La premi\`ere \'etape est de s\'eparer les parties quadratiques et quartiques de l'\'energie pour \'ecrire (on prendra $g=t=1$ dans toute la suite)
\begin{equation}\label{eq:NL Gibbs split}
\mu (du) = \underbrace{\frac{1}{z_r} \exp\left( - \frac{1}{2} \iint_{\R^d\times \R^d} |u(x)|^2 w (x-y) |u(y)|^2 dx dy \right)}_{\mbox{poids non lin\'eaire}}  \underbrace{\frac{1}{z_{0}} \exp\left( - \left\langle u | -\Delta + V | u \right\rangle \right) du}_{\mbox{mesure gaussienne}} 
\end{equation}
avec $z_r$ une fonction de partition relative. 

\subsection{Mesures (gaussiennes) sans interaction}\label{sec:measures gauss}

Le point est de voir la mesure compl\`ete comme absolument continue par rapport \`a la mesure gaussienne, que l'on peut d\'efinir de mani\`ere naturelle. Pour simplifier la suite de l'exposition on se limitera \`a des op\'erateurs de Schr\"odinger \`a une particule de forme 
\begin{equation}\label{eq:op Schro}
 h = -\Delta + V = - \Delta + |x| ^s = \sum_{j=1} ^\infty \lambda_j |u_j \rangle \langle u_j | 
\end{equation}
avec $s >1$ et on utilisera les espaces de type Sobolev associ\'es \`a la d\'ecomposition spectrale ci-dessus
$$ 
\gH^{t} = D (h^{t/2}) = \left\{ u = \sum_{j=1} ^{\infty} \alpha_j u_j \: | \: \sum \lambda_j |\alpha_j| ^2 < \infty \right\}.
$$
A $t=0$ on a l'espace $L^2 (\R^d)$ usuel, \`a $t=1$ l'espace d'\'energie, mais on sera aussi amen\'e \`a travailler dans des espaces \`a $t<0$ qui sont a priori (et dans beaucoup de cas, a posteriori aussi) des espaces de distributions plut\^ot que de fonctions.

\begin{lemma}[\textbf{Mesure gaussienne}]\label{lem:gaussienne}\mbox{}\\
Supposons que 
\begin{equation}\label{eq:p trace}
 \tr ( h ^{-p}) = \sum_{j=1} ^\infty \lambda_j ^{-p} < \infty. 
\end{equation}
Soit, pour $K$ un entier positif,  
$$ \mu_{0,K} (du) := \prod_{j=1} ^K \frac{\lambda_j}{\pi} \exp\left( - \lambda_j \left| \langle u | u_j\rangle \right| ^2 \right) d \langle u | u_j \rangle $$
avec $d \langle u | u_j \rangle$ la mesure de Lebesgue usuelle sur $\C \equiv \R ^2$. 
La suite de mesures $(\mu_{0,K})_K$ est tendue sur $\gH^{1-p}$ et il existe une unique mesure de probabilit\'e sur $\gH^{1-p}$ telle que $\mu_{0,K}$ soit la projection cylindrique de $\mu_0$ sur l'espace engendr\'e par $u_1,\ldots,u_K$.
\end{lemma}
 
On appelle $\mu_0$ la \emph{mesure gaussienne} de covariance $h^{-1}$. On voit qu'elle n'est a priori pas d\'efinie sur des espaces fonctionnels tr\`es r\'eguliers, puisque $p>0$ et m\^eme assez souvent $p>1$. Il est en fait connu que les conditions de support trouv\'ees ci-dessus ne peuvent s'am\'eliorer (th\'eor\`eme de Fernique): si $p$ est le meilleur (plus petit) nombre tel que~\eqref{eq:p trace} ait lieu, $\mu_0$ est support\'ee sur $\gH ^{1-p}$ mais assigne une mesure $0$ \`a $\gH ^{1-p'}$ pour tout $p'<p$. En particulier, un fait quelque peu non intuitif est que, bien que la mesure soit de forme ``exponentielle de moins l'\'energie'', l'\'energie est en fait presque partout infinie sur son support, puisque la mesure de $\gH^1$ est nulle. L'explication intuitive est que, dans un espace de dimenion infinie, il y a tellement de place qu'il est favorable de rendre l'\'energie infinie en chargeant de plus gros espaces fonctionnels, ceci favorisant l'entropie.

\begin{remark}[Valeurs de $p$ ayant une pertinence particuli\`ere]
\flushleft Par la suite on sera plus particul\`erement int\'eress\'es par deux cas du lemme ci-dessus:

\smallskip

\noindent \textbf{Cas \`a trace}, $p\leq 1$ et en particulier, $h^{-1}$ est un op\'erateur \`a trace (ses valeurs propres forment une suite sommable). C'est le cas le plus simple. En particulier, $\mu_0$ charge (un sous-espace de) $L^2 (\R^d)$.  

\smallskip

\noindent \textbf{Cas Hilbert-Schmidt}, $ 1 < p \leq 2$ et en particulier $h^{-1}$ est un op\'erateur Hilbert-Schmidt (ses valeurs propres forment une suite de carr\'e sommable). Ici l'analyse devient beaucoup plus complexe: le support de $\mu_0$ ne contient a priori que des distributions et pas de fonctions. 

\smallskip

\hfill $\diamond$
\end{remark}

Il est assez facile de d\'ecider dans quel cas on se trouve en utilisant des in\'egalit\'es semi-classiques de type Lieb-Thirring (voir~\cite[Chapitre~4]{LieSei-09} et r\'ef\'erences cit\'ees). Dans le cas~\eqref{eq:op Schro} on obtient~\eqref{eq:p trace} pour tout 
$$ p > \frac{d}{2} + \frac{d}{s}$$
et donc:

\begin{example}[Mesures associ\'ees \`a des op\'erateurs de Schr\"odinger]\label{exe:s,p,d}
\flushleft Le support de la mesure gaussienne d\'efinie ci-dessus, dans le cas~\eqref{eq:op Schro} d\'epend assez fortement de $d$ la dimension d'espace et de $s$ l'exposant fixant la croissance du potentiel \`a l'infini. On inclut formellement le cas de $h= -\Delta + 1$ sur un ouvert born\'e avec conditions de bord \`a choisir\footnote{$-\Delta +1$ plut\^ot que $-\Delta$ pour \'eviter un mode d'\'energie nulle et pouvoir d\'efinir $h^{-p}$ quelle que soit la condition de bord.} en prenant $s=\infty$.

\smallskip

\noindent\textbf{Une dimension d'espace.} Le meilleur $p$ possible vaut $1/2 + 1/s$. On est donc dans le cas \`a trace pour $s>2$. La distinction avec le cas Hilbert-Schmidt n'est cependant pas tr\`es pertinente en 1D, voir plus bas et~\cite{LewNamRou-17}.
 
 \smallskip

\noindent \textbf{Deux dimensions d'espace.} Le meilleur $p$ possible vaut $1 + 2/s$. On est donc \emph{jamais} dans le cas \`a trace. On reste dans le cas Hilbert-Schmidt pour $s>2$.

\smallskip

\noindent \textbf{Trois dimensions d'espace.} Le meilleur $p$ possible vaut $3/2 + 3/s$. On est \`a nouveau \emph{jamais} dans le cas \`a trace. On reste dans le cas Hilbert-Schmidt pour $s>6$.
 
\smallskip

\hfill $\diamond$
\end{example}

\subsection{Mesures (non-lin\'eaires) avec interaction}\label{sec:meas non lin}

A ce stade nous avons d\'efini pr\'ecis\'ement le deuxi\`eme facteur du membre de droite de~\eqref{eq:NL Gibbs split}. Il nous reste \`a g\'erer le premier facteur, et les difficult\'es sont apparentes. Contentons-nous pour simplifier de chercher \`a d\'efinir l'interaction pour un potentiel $w$ r\'egulier. M\^eme dans ce cas, il n'est pas toujours clair de donner un sens \`a l'argument de l'exponentielle. A part dans le cas \`a trace, $u$ est presque s\^urement une distribution. Que signifient les produits dans ce cas ?  

Avant de r\'esoudre cette difficult\'e, d\'ebarassons-nous du cas \`a trace, beaucoup plus simple (vu l'Exemple~\ref{exe:s,p,d} le cas \`a trace est assez sp\'ecifiquement 1D). 

\begin{lemma}[\textbf{Mesure avec interactions, cas \`a trace}]\label{lem:NL Gibbs trace}\mbox{}\\
Supposons que~\eqref{eq:p trace} ait lieu avec $p=1$. Soit alors $\mu_0$ la mesure gaussienne d\'efinie par le Lemme~\ref{lem:gaussienne} et $w$ un potentiel d'interaction de forme
$$ 0 \leq w \in L^{\infty}(\R) + c \delta_0$$
avec $\delta_0$ la masse de Dirac \`a l'origine. La fonction 
\begin{equation}\label{eq:int 1D}
 u\mapsto \cD [|u|^2] := \frac{1}{2} \iint_{\R\times \R} |u(x)| ^2 w (x-y) |u(y)| ^2 dx dy \geq 0 
\end{equation}
est dans $L^1 (d\mu_0)$. Cons\'equemment
\begin{equation}\label{eq:NL Gibbs 1D} 
\mu (du) := \frac{1}{z_r} \exp\left( -\cD [|u|^2]  \right) d\mu_0(u)
\end{equation}
fait sens comme mesure de probabilit\'e.
\end{lemma}

Remarquons que en 1D,~\eqref{eq:p trace} peut cesser d'avoir lieu avec $p=1$ sans que de grosses difficult\'es conceptuelles ne soient associ\'ees~\cite{LewNamRou-17}. La raison en est qu'il est fort possible que~\eqref{eq:int 1D} continue de faire sens, bien que $\mu_0$ ne vive pas sur $\gH_0 = L ^2 (\R)$. En fait, la mesure continue de vivre sur $L^4 (\R)$ tant que $s>1$, ce qui permet de donner un sens \`a l'interaction cf~\cite{BurThoTzv-09,BurThoTzv-10,LewNamRou-17}. 

\medskip

Quittons maintenant le cas \`a trace et ses satellites. Le probl\`eme pour d\'efinir le premier facteur du membre de droite de~\eqref{eq:NL Gibbs split} est que, $u$ \'etant $\mu_0$-presque s\^urement une distribution, il n'y a aucune chance que~\eqref{eq:int 1D} fasse sens dans ce cas. C'est une difficult\'e bien connue dans tous les domaines qui utilisent de telles mesures: th\'eorie constructive des champs, th\'eorie de Cauchy probabiliste pour l'\'equation NLS, \'etude en temps longs de l'\'equation de la chaleur stochastique non lin\'eaire (voir r\'ef\'erences cit\'ees plus haut) ...  

La solution pour d\'efinir une mesure bas\'ee sur la d\'ecomposition~\eqref{eq:NL Gibbs split} est de \emph{renormaliser} l'interaction en soustrayant des contre-termes formellement infinis. Plus pr\'ecis\'ement, on projette la mesure en dimension finie, et on s'assure que les contre-termes divergent de la m\^eme fa\c{c}on que le terme principal lorsque la dimension tend vers l'infini, afin d'avoir un objet final bien d\'efini. Trouver les contre-termes appropri\'es et contr\^oler cette proc\'edure constitue une th\'eorie en soi, surtout pour des interactions singuli\`eres. Mentionnons pour m\'emoire le travail intense d\'evou\'e \`a la d\'efinition de la mesure dite $\Phi^4_3$ (interactions ponctuelles $\Phi^4$ en trois dimensions d'espace) et tournons-nous vers les cas, beaucoup plus simples, qui vont nous int\'eresser ici. La lectrice a peut-\^etre d\'ej\`a rencontr\'e la construction de $\Phi_2 ^4$, essentiellement due \`a Nelson~\cite{Nelson-66}, et a peut-\^etre lu dans~\cite{Simon-Nelson} que c'\'etait la plus simple des renormalisations possibles. Ce dernier point est en fait erron\'e, voici la plus simple\footnote{La tournure un peu ironique de cette phrase dissimule le fait que cette construction est sans int\'er\^et pour la th\'eorie constructive des champs quantiques, puisqu'elle traite d'interactions non-locales. De plus, les experts du domaine la connaissent s\^urement, bien que je ne l'ai pas vue imprim\'ee avant~\cite{Bourgain-97}.} des renormalisations possibles:

\begin{lemma}[\textbf{Mesure avec interactions, cas Hilbert-Schmidt.}]\label{lem:NL Gibbs HS}

\flushleft Supposons que~\eqref{eq:p trace} ait lieu avec $p=2$. Soit alors $\mu_0$ la mesure gaussienne d\'efinie par le Lemme~\ref{lem:gaussienne} et, pour $f$ une fonction de $L^1 (d\mu_0)$, la valeur moyenne associ\'ee: 
$$ \langle f(u) \rangle_{\mu_0} := \int f(u) d\mu_0 (u).$$
Soit $w$ un potentiel d'interaction dont la transform\'ee de Fourier $\hat{w}$ satisfait
$$ 0 \leq \hat{w} \in L^{1}(\R^d).$$
Soit $P_K$ le projecteur orthogonal sur les $K$ premiers modes propres de $h$. 
La fonction 
\begin{multline}\label{eq:int 2D}
 u\mapsto \cD^R_K [|u|^2] := \\
 \frac{1}{2} \iint_{\R^d\times \R^d} \left(|P_K u(x)| ^2 - \langle |P_K u(x)| ^2 \rangle_{\mu_0} \right) w (x-y) \left(|P_K u(y)| ^2 - \langle |P_K u(y)| ^2 \rangle_{\mu_0} \right) dx dy \\
 \geq 0 
\end{multline}
d\'efinit une suite de Cauchy dans $L^1 (d\mu_0)$. Cons\'equemment elle converge pour $K\to \infty$ dans $L^1 (d\mu_0)$ vers une fonction $\cD^R [|u|^2]$ et 
\begin{equation}\label{eq:NL Gibbs 2D} 
\mu (du) := \frac{1}{z_r} \exp\left( -\cD^R [|u|^2]  \right) d\mu_0(u)
\end{equation}
fait sens comme mesure de probabilit\'e.
\end{lemma}

Quelques commentaires:

\smallskip

\noindent(1) On retrouve bien l'esprit de la renormalisation. Tout est d'abord projet\'e en dimension finie en utilisant $P_K$, des termes sont soustraits dans~\eqref{eq:int 1D} pour obtenir une limite bien d\'efinie. L'intuition est ici que $|u(x)|^2$ est infini $\mu_0$ presque s\^urement, mais que cet infini ``est le m\^eme'' $\mu_0$ presque s\^urement. Il suffit donc de soustraire sa moyenne \`a chaque occurence de $|u(x)| ^2$ dans la d\'efinition de l'interaction.

\smallskip

\noindent(2) Le contr\^ole de la proc\'edure est beaucoup plus simple que dans les cas embl\'ematiques tr\`es \'etudi\'es o\`u $w$ est une masse de Dirac. Un potentiel plus r\'egulier facilite la t\^ache et constitue d\'ej\`a un cas int\'eressant dans le cadre qui nous occupe (on r\'ealise en fait ici un ordre de Wick partiel, au lieu de l'ordre de Wick complet n\'ecessit\'e par $\Phi_2 ^4$). Remarquons qu'en supposant $\hat{w} \geq 0$, l'interaction reste positive apr\`es la renormalisation, ce qui est une simplification de plus par rapport \`a la proc\'edure usuelle.

\smallskip

\noindent(3) Notons que cette proc\'edure fonctionne, pour un potentiel $w$ relativement r\'egulier, dans tout le cas Hilbert-Schmidt, donc en 2D et en 3D, cf Exemple~\ref{exe:s,p,d}. Nous serons amen\'es \`a plus de restrictions dans le cas quantique ci-dessous, nous limitant \`a deux dimensions d'espace.

\smallskip

Finalement, une derni\`ere remarque m\'erite d'\^etre isol\'ee, comme plus importante pour la suite de cet expos\'e: 

\begin{remark}[Renormalisation : termes direct et d'\'echange]\label{rem:direct exchange}

\flushleft Il est instructif de calculer formellement l'esp\'erance des interactions renormalis\'ees et non-renormalis\'ees dans la mesure de Gibbs libre $\mu_0$. En utilisant le th\'eor\`eme de Wick satisfait par la mesure gaussienne on trouve, pour la version non-renormalis\'ee
\begin{multline}\label{eq:esp int non renor}
\int \left(\iint_{\R^d \times \R^d} |u(x)| ^2 w(x-y) |u(y)|^2 dx dy \right) d\mu_0 (u) \\ = \iint_{\R^d \times \R^d} G(x,x) w(x-y) G(y,y) dx dy + \iint_{\R^d \times \R^d} |G(x,y)| ^2 w(x-y) dx dy 
\end{multline}
et pour la version renormalis\'ee
\begin{align}\label{eq:esp int renor}
\int \cD^R [|u|^2] d\mu_0 (u) = \iint_{\R^d \times \R^d} |G(x,y)| ^2 w(x-y) dx dy 
\end{align}
avec $G(x,y)$ la fonction de Green du Hamiltonien $h$ (noyau int\'egral de $h^{-1}$):
$$ \left(-\Delta_x + V (x) \right) G(x,y) = \delta_{x=y}.$$
Bien s\^ur le premier terme de~\eqref{eq:esp int non renor} est purement formel puisque dans le cas qui nous int\'eresse la densit\'e de mati\`ere 
$$ \rho (x):= G (x,x) \equiv + \infty$$
$G(x,y)$ ayant, en dimensions $2$ et sup\'erieures une singularit\'e au moins logarithmique pour $x\sim y$. Par contre, dans le cas Hilbert-Schmidt $p=2$, $(x,y)\mapsto G(x,y)$ est au moins $L^2$ et donc le deuxi\`eme terme de~\eqref{eq:esp int non renor} fait sens. Ces deux termes portent les noms de \emph{terme direct} et \emph{terme d'\'echange}, le premier correspondant \`a l'interaction classique, de type champ moyen, de la densit\'e de mati\`ere correspondant \`a $\mu_0$ avec elle-m\^eme. On voit que l'effet de la renormalisation dans~\eqref{eq:esp int renor} est de d\'etruire ce terme, infini (ou plus pr\'ecis\'ement, divergent dans l'ultraviolet), et de ne garder que le terme d'\'echange, fini (convergent dans l'ultraviolet).  \hfill$\diamond$
\end{remark}

\section{R\'esultats principaux \`a temp\'erature positive}

Nous nous attelons maintenant \`a rendre rigoureuses les connections informelles esquiss\'ees \`a la Section~\ref{sec:can}. Les d\'efinitions des objets limites ayant \'et\'ees rappell\'ees ci-dessus, il nous faut maintenant \^etre plus pr\'ecis sur les objets de d\'epart et sur le r\'egime de param\`etres permettant de faire la connection. Ceci nous occupera dans la Section~\ref{sec:grand can}. Dans les Sections~\ref{sec:main trace} et~\ref{sec:main HS} respectivement nous pourrons ensuite \'enoncer les r\'esultats obtenus dans~\cite{LewNamRou-14d,FroKnoSchSoh-16} et~\cite{LewNamRou-18c}, portant sur le cas \`a trace et le cas Hilbert-Schmidt respectivement (voir \'egalement~\cite{FroKnoSchSoh-17} pour des r\'esultats dans un cadre d\'ependant du temps). La nouveaut\'e principale des r\'esultats que nous allons exposer est qu'ils traitent de m\'ecanique statistique quantique avec un espace \`a une particule de dimension infinie. En dimension finie les choses sont beaucoup plus simples, voir~\cite{Knowles-thesis,Gottlieb-05} et~\cite[Appendice~B]{Rougerie-spartacus}.

Faute de place nous ne discuterons pas de la lit\'erature physique consacr\'ee \`a des questions connexes, mais voir~\cite{ArnMoo-01,BayBlaiHolLalVau-99,BayBlaiHolLalVau-01,HolBay-03,KasProSvi-01,BisDavSimBla-09,GioCarCas-07,HolCheKra-08,HolKra-08,ProSvi-02,ProRueSvi-01,SimDavBlak-08}.

\subsection{Ensemble grand-canonique}\label{sec:grand can}

A la Section~\ref{sec:can} nous avons d\'ecrit le formalisme appropri\'e pour un nombre fix\'e de bosons \`a temp\'erature positive. Cependant, les objets limites naturels d\'efinis \`a la Section~\ref{sec:meas non lin} n'ont pas un nombre de particules fixe: $\int_{\R^d} |u|^2$ est une variable al\'eatoire dans ce contexte. Il est d\`es lors clair qu'il nous faut partir d'un cadre $N$-corps quantique permettant des fluctuations du nombre de particules: l'ensemble grand-canonique. D'un point de vue de physique statistique ce changement de perspective n'est pas tr\`es important: on s'attend \`a ce que les ensembles canoniques et grand-canoniques donnent des r\'eponses similaires dans les situations et limites d'int\'er\^et physique. 

\medskip

\noindent\textbf{Espace de Fock, seconde quantification.} On rassemble tous les espaces \`a $N$ particules dans une somme directe, l'espace de Fock bosonique
\begin{equation}\label{eq:Fock}
\gF := \C \oplus \gH \oplus \gH_2 \oplus \ldots \oplus \gH_n \oplus \ldots
\end{equation}
avec comme ci-dessus 
$$ \gH_n = L^2_{\rm sym} (\R ^{dn}).$$
Un \'etat (grand canonique) $\Gamma$ sur l'espace de Fock est un op\'erateur auto-adjoint positif de trace $1$ sur $\cF $. Si il commute avec le nombre de particules (qui devient un op\'erateur dans ce contexte)
\begin{equation}\label{eq:grand can nombre}
 \cN := \bigoplus_n n \1_{\gH_n} 
\end{equation}
on peut l'\'ecrire 
\begin{equation}\label{eq:etat diagonal}
 \Gamma = \Gamma_0 \oplus \Gamma_1 \oplus \ldots \oplus \Gamma_n \oplus \ldots
\end{equation}
i.e comme une superposition d'op\'erateurs positifs \`a nombre de particules fixes, li\'es par la normalisation 
$$ \tr_{\gF} \Gamma = \sum_n \tr_{\gH_n} \Gamma_n = 1.$$
Etant donn\'e un op\'erateur \`a $k$ particules $A_k$ agissant sur $\gH_k$ on peut \'etendre son action \`a l'espace de Fock entier (seconde quantification) en posant 
\begin{equation}\label{eq:second quant}
 \mathbb{A}_k := \bigoplus_{n\geq k} \, \sum_{1\leq i_1 < \ldots < i_k \leq n} (A_k)_{i_1,\ldots,i_k}  
\end{equation}
o\`u $(A_k)_{i_1,\ldots,i_k}$ est l'op\'erateur $A_k$ agissant sur les variables $i_1,\ldots,i_k$ de l'espace $\gH_n$. De mani\`ere duale, on s'int\'eresse la plupart du temps \`a l'action de tels op\'erateurs \`a nombre de particules donn\'e sur un \'etat grand canonique. Ceci d\'efinit la notion de matrice de densit\'e r\'eduite $\Gamma ^{(k)}$, agissant sur l'espace \`a $k$ particules $\gH_k$:
$$ \tr_{\gH_k} \left( A_k \Gamma^{(k)}\right) = \tr_{\gF} \left( \mathbb{A}_k \Gamma \right)$$
pour tout op\'erateur born\'e $A_k$. Si $\Gamma$ est (comme dans tous les cas qui nous int\'eresseront) de forme~\eqref{eq:etat diagonal} on a alors en termes de traces partielles sur les deg\'es de libert\'e exc\'edentaires 
$$ \Gamma ^{(k)} = \sum_{n\geq k} \tr_{k+1 \to n} \Gamma_n.$$
Les physiciens pr\'ef\`erent (avec raison(s)) une d\'efinition \'equivalente\footnote{La lectrice familiaris\'ee verra l'\'equivalence facilement, et nous pardonnera d'\'epargner un autre \'el\'ement de formalisme au lecteur non familiaris\'e. } des matrices de densit\'e r\'eduites, utilisant les op\'erateurs de cr\'eation/annihiliation.

\medskip

\noindent\textbf{Hamiltonien, \'energie libre et \'etat de Gibbs.} Nous \'etendons de mani\`ere naturelle le formalisme \'evoqu\'e \`a la Section~\ref{sec:can} au cadre de l'espace de Fock. La contrainte de nombre de particules fix\'e est remplac\'ee par l'inclusion d'un potential chimique $\nu \in \R$ servant \`a r\'egler le poids de l'op\'erateur nombre de particules~\eqref{eq:grand can nombre}. L'\'energie libre d'un \'etat $\Gamma$ sur $\gF$ est donn\'ee par 
\begin{equation}\label{eq:grand can free func}
\cF_\lambda [\Gamma] = \tr_{\gF} \left( \left( \mathbb{H}_\lambda - \nu \cN + E_0 \right) \Gamma \right) + T \tr_{\gF} \left( \Gamma \log \Gamma \right).  
\end{equation}
Le second terme est l'oppos\'e de l'entropie, multipli\'e par la temp\'erature $T$ comme il se doit. Le premier terme contient 

\smallskip

\noindent$\bullet$ l'\'energie proprement dite, encod\'ee par le Hamiltonien $\bH_\lambda$. Ce dernier est l'extension naturelle de~\eqref{eq:N body Schro} \`a l'espace de Fock: 
$$ \bH_\lambda:= \bigoplus_n H_{n,\lambda}$$ 
avec $\lambda \geq 0$ une constante de couplage. 

\smallskip

\noindent$\bullet$ l'ajustement du nombre de particules au moyen d'un potentiel chimique $\nu >0$, ce qui remplace sa d\'etermination exacte et fixe dans le cadre canonique.

\smallskip

\noindent$\bullet$ une \'energie de r\'ef\'erence $E_0$, utile uniquement pour faciliter certains \'enonc\'es ci-dessous. 

\smallskip

Nous nous int\'eresserons \`a la minimisation de l'\'energie libre parmi tous les \'etats grands-canoniques:
\begin{equation}\label{eq:grand can free ener}
F_\lambda = \min \left\{ \cF_\lambda [\Gamma] \,\big|\, \Gamma \mbox{ op\'erateur \`a trace positif sur } \gF, \tr_{\gF} (\Gamma) = 1 \right\}. 
\end{equation}
La minimisation est ``explicite'' et donne l'\'etat de Gibbs comme unique minimiseur:
\begin{equation}\label{eq:grand can Gibbs}
 \Gamma_\lambda = \frac{1}{\cZ_\lambda} \exp\left( - \frac{1}{T} \bH_\lambda \right)
\end{equation}
avec la fonction de partition 
$$ \cZ_\lambda = \tr_{\gF} \left( \exp\left( - \frac{1}{T} \bH_\lambda \right)\right)$$
satisfaisant la relation 
$$ F_\lambda = - T \log \cZ_\lambda.$$
Nous nous int\'eresserons par la suite au comportement asymptotique de l'\'energie libre $F_\lambda$ et des (matrices de densit\'e r\'eduites des) \'etats de Gibbs associ\'es, dans un certain r\'egime $T\to \infty$ impliquant d'ajuster pr\'ecis\'ement les param\`etres $\lambda, \nu, E_0$ pour obtenir une limite non triviale.

\subsection{Une dimension d'espace, pas de renormalisation}\label{sec:main trace}

Consid\'erons d'abord la limite des \'etats de Gibbs grand-canoniques ci-dessus dans le cas \`a trace, c'est-\`a-dire essentiellement en une dimension d'espace, et pour un potentiel ext\'erieur assez confinant. La relation entre les objets quantique et classique (introduits au Lemme~\ref{lem:NL Gibbs trace}) est alors assez imm\'ediate (\`a \'enoncer, pas \`a prouver !). Le r\'esultat suivant a \'et\'e prouv\'e dans~\cite{LewNamRou-14d}, puis dans~\cite{FroKnoSchSoh-16} avec une preuve diff\'erente: 

\newpage

\begin{theorem}[\textbf{Limite de champ moyen \`a temp\'erature positive en 1D}]\label{thm:main trace}
\flushleft Soit $\Gamma_\lambda$ l'\'etat de Gibbs grand-canonique associ\'e \`a~\eqref{eq:N body Schro} avec $d=1$ et 
$$ h = - \Delta + V = - \partial_x ^2 + |x| ^s, s>2.$$
On suppose que le potentiel d'interaction $w$ satisfait les hypoth\`eses du Lemme~\ref{lem:NL Gibbs trace}. On choisit les param\`etres de~\eqref{eq:grand can free func} comme suit: 
$$ T \to \infty, \quad \lambda = \frac{1}{T}, \quad \nu < \lambda_1 \mbox{ fixe}, \quad E_0 = 0$$
avec $\lambda_1$ la premi\`ere valeur propre de $h$. 

Soit $\mu$ la mesure de Gibbs non-lin\'eaire d\'efinie au Lemme~\ref{lem:NL Gibbs trace}, avec cette fois 
$$ h = - \partial_x ^2 + |x| ^s - \nu $$
et $\mu_0$ la mesure gaussienne associ\'ee \`a cet op\'erateur. On a alors, dans la limite $T\to \infty$, 
\begin{enumerate}
 \item Asymptotique de l'\'energie libre: avec $F_\lambda$ l'\'energie libre associ\'ee \`a~\eqref{eq:grand can Gibbs} et $z_r$ la fonction de partition relative classique d\'efinie au Lemme~\ref{lem:NL Gibbs trace}
 \begin{equation}\label{eq:main 1D ener libre}
\frac{F_\lambda - F_0}{T} \underset{T\to \infty}{\longrightarrow} - \log z_r.  
 \end{equation}
\item Asymptotique des matrices de densit\'es r\'eduites: soit $\Gamma_\lambda ^{(k)}$ la $k$-\`eme matrice de densit\'e r\'eduite de l'\'etat de Gibbs~\eqref{eq:grand can Gibbs}. On a, pour tout $k\in \N$, 
\begin{equation}\label{eq:main 1D states}
\tr_{\gH_k} \left| \frac{k!}{T^k} \Gamma_\lambda ^{(k)} - \int |u ^{\otimes k} \rangle \langle u ^{\otimes k} | d\mu(u) \right| \underset{T\to \infty}{\longrightarrow} 0.  
\end{equation}
\end{enumerate}

\end{theorem}

Quelques commentaires:

\smallskip

\noindent(1) La limite $T\to \infty$ (avec le choix de potentiel chimique $\nu$ fixe ci-dessus) permet de mimer une limite de champ moyen usuelle, canonique, o\`u l'on ferait tendre le nombre de particules vers l'infini. Dans le cas \`a trace qui nous int\'eresse pr\'esentement, le nombre de particules est proportionel \`a $T$ (comme montr\'e a posteriori par~\eqref{eq:main 1D states}), et il alors naturel de choisir une constante de couplage $\lambda \propto T^{-1}$ pour balancer l'effet des interactions et celui du Hamiltonien \`a une particule. L'\'energie de r\'ef\'erence $E_0$ ne joue pas de r\^ole ici.

\smallskip

\noindent(2) Il est assez naturel que l'asymptotique de l'\'energie libre~\eqref{eq:main 1D ener libre} concerne la diff\'erence entre l'\'energie avec interactions et celle sans interactions ($\lambda = 0$). Les deux quantit\'es prises s\'epar\'ement divergent beaucoup plus vite que $T$.

 \smallskip

\noindent(3) On voit ici le ``sens pr\'ecis'' qu'il faut donner \`a la convergence d'\'etats quantiques dont le nombre de particules augmente: convergence des matrices de densit\'e r\'eduites.
 
 \smallskip

\noindent(4) La convergence des matrices de densit\'e r\'eduites~\eqref{eq:main 1D states} a lieu en norme de trace, la topologie la plus naturelle. Elle montre que l'on peut penser \`a l'\'etat de Gibbs originel comme \`a une superposition statistique 
 $$ \Gamma_\lambda \approx \int | \xi \left( \sqrt{T} u \right) \rangle \langle \xi \left( \sqrt{T} u \right)| d\mu (u) $$
 d'\'etats coh\'erents 
 $$ \xi(v) = e^{-\norm{v} ^2 }\bigoplus_n \frac{v ^{\otimes n}}{n!}$$
qui jouent, dans le contexte grand-canonique, le m\^eme r\^ole que les \'etats factoris\'es $v ^{\otimes N}$ dans le contexte canonique.
 
 \smallskip

\noindent(5) Ce r\'esultat a une extension au cas o\`u $s>1$. A nouveau, aucune renormalisation n'est n\'ecessaire dans ce cas, bien que cela soit moins directement visible. Voir~\cite{LewNamRou-17} et sa liste de r\'ef\'erences.

\subsection{Deux dimensions d'espace, renormalisation}\label{sec:main HS}

Nous \'enon\c{c}ons ici une partie des r\'esultats principaux de~\cite{LewNamRou-18c}, un r\'esum\'e plus succinct et partiel se trouvant dans~\cite{LewNamRou-18b}. Ils couvrent une partie des mesures d\'efinies au Lemme~\ref{lem:NL Gibbs HS}. La conjecture naturelle est que les r\'esultats ci-dessous s'\'etendent dans tous les cas couverts par le Lemme~\ref{lem:NL Gibbs HS} (i.e. $1<p<2$) mais ceci reste un probl\`eme ouvert. Au prix d'une modification quelque peu ad-hoc de l'\'etat de Gibbs de d\'epart~\eqref{eq:grand can Gibbs}, cette extension est contenue dans~\cite{FroKnoSchSoh-16} (qui pr\'ec\`ede chronologiquement~\cite{LewNamRou-18c}). 

Il y a une diff\'erence majeure dans la formulation des r\'esultats pour les cas \`a trace (Section~\ref{sec:main trace} ci-dessus) et Hibert-Schmidt (pr\'esente section), qui refl\`ete le besoin de renormaliser l'objet limite discut\'e pr\'ec\'edement. On a vu \`a la Remarque~\ref{rem:direct exchange} que la renormalisation revenait \`a ignorer le terme direct de l'\'energie d'interaction. Or ce terme a bel et bien un pendant dans l'\'energie libre quantique~\eqref{eq:grand can free func}. Il se trouve que, dans les cas qui nous occupent ici, et contrairement au cas \`a trace de la Section~\ref{sec:main trace}, ce terme a un poids si important qu'il modifie la mesure gaussienne de r\'ef\'erence. La mesure appropri\'ee est en fait singuli\`ere par rapport \`a la mesure non-interagissante qui serait le choix na\"if de mesure de r\'ef\'erence. Notre premi\`ere t\^ache est donc, en suivant~\cite{FroKnoSchSoh-16}, de trouver la bonne mesure de r\'ef\'erence en r\'esolvant un probl\`eme non lin\'eaire de type champ moyen. 

\medskip

\noindent\textbf{Etat quasi-libre de r\'ef\'erence.} Une premi\`ere approximation commun\'ement utilis\'ee pour le probl\`eme de minimisation~\eqref{eq:grand can free ener} est de restreindre la minimisation aux \'etats quasi-libres, plus pr\'ecis\'ement \`a ceux de la forme 
\begin{equation}\label{eq:quasi libre}
 \Gamma = \frac{\exp\left(- T^{-1} \dG (h) \right)}{\tr_{\gF} \left( \exp\left(- T^{-1} \dG (h) \right) \right)} 
\end{equation}
avec $h$ un Hamiltonien \`a une particule et $\dG (h)$ sa seconde quantification donn\'ee par~\eqref{eq:second quant}. Trouver le bon \'etat quasi-libre de r\'ef\'erence comprend les \'etapes suivantes: 

\smallskip

\noindent(1) Minimiser l'\'energie libre~\eqref{eq:grand can free func} restreinte aux \'etats de forme~\eqref{eq:quasi libre} et obtenir un minimiseur $\tilde{\Gamma}_0$.
 
 \smallskip

\noindent(2) Associer \`a $\tilde{\Gamma}_0$ le Hamiltonien de champ moyen $\tilde{h}$ tel que 
 $$ \tilde{\Gamma}_0 = \frac{\exp\left(- T^{-1} \dG (\tilde{h}) \right)}{\tr_{\gF} \left( \exp\left(- T^{-1} \dG (\tilde{h}) \right) \right)}.$$
La relation entre $\tilde{\Gamma}_0$ et $\tilde{h}$ est non-lin\'eaire.

\smallskip

\noindent(3) Observer que, si on ajuste correctement le potentiel chimique $\nu$ apparaissant dans~\eqref{eq:grand can free func}, le Hamiltonien \`a une particule $\tilde{h}$ (qui d\'epend de tous les param\`etres du probl\`eme) a une limite $h_\infty$ quand $T\to \infty$ (en un certain sens). 

\smallskip

\noindent(4) La mesure gaussienne appropri\'ee pour d\'efinir les objets limites du probl\`eme quantique complet est celle associ\'ee \`a $h_{\infty}$. 

\smallskip

Ce programme a \'et\'e men\'e \`a bien dans~\cite{FroKnoSchSoh-16}, et le th\'eor\`eme qui suit est une reformulation de certains de ces r\'esultats. Nous faisons une petite entorse simplificatrice au point $(1)$ ci-dessus: on minimisera parmi les \'etats quasi-libres l'\'energie libre dite de Hartree (par opposition \`a Hartree-Fock), \emph{sans terme d'\'echange}. Techniquement cela facilite l'\'etude, conceptuellement cela n'a pas d'importance: le terme d'\'echange sera pris en compte plus tard.

Une autre fa\c{c}on de pr\'esenter le programme ci-dessus est sous la forme de la recherche d'un contre-terme appropri\'e compensant la divergence de l'\'energie d'interaction directe. Le point le plus d\'elicat est de montrer que (comme d\'ecrit ci-dessus) ce contre-terme peut \^etre fourni en ajustant seulement le potentiel chimique $\nu$. Nous renvoyons aux articles originaux pour plus de commentaires et passsons maintenant \`a l'\'enonc\'e du 

\begin{theorem}[\textbf{Etat quasi-libre de r\'ef\'erence}]\label{thm:contre}\mbox{}\\
Consid\'erons la fonctionnelle d'\'energie libre de Hartree(-Fock r\'eduite), d\'efinie pour un op\'erateur \`a trace positif $\gamma$ sur $\gH$:
\begin{multline}\label{eq:rHF}
\cF^{\rm rH}[\gamma]=\Tr\left[(-\Delta+V-\nu)\gamma\right]+\frac{\lambda}{2} \iint \gamma(x;x) w(x-y) \gamma(y;y)\, dx\, dy  \\
-T\tr\left[(1+\gamma)\log(1+\gamma)-\gamma\log\gamma\right].
\end{multline}
Prenons $V(x) = |x| ^s$ avec $s$ assez grand, et $w\in L^1 (\R ^d)$ \`a transform\'ee de Fourier $\hat{w}\in L^1 (\R ^d)$ positive.  Les faits suivants sont vrais: 
\begin{enumerate}
 \item La fonctionnelle~\eqref{eq:rHF} a un unique minimiseur $\gamma^{\rm rH}$, qui est la premi\`ere matrice de densit\'e r\'eduite de l'\'etat quasi-libre 
\begin{equation}\label{eq:quasi-libre ref}
 \tilde{\Gamma}_0 = \frac{\exp\left(- T^{-1} \dG (\tilde{h}) \right)}{\tr_{\gF} \left( \exp\left(- T^{-1} \dG (\tilde{h}) \right) \right)} 
\end{equation}
avec $\tilde{h} = -\Delta + V_T$ et
\begin{equation}\label{eq:ren pot}
V_T = \lambda \gamma^{\rm rHF} (x;x) * w +V - \nu. 
\end{equation}
\item Prenons une constante $\kappa >0$ et fixons $\nu$ comme suit:
\begin{equation}\label{eq:rhoOkappa}
\nu := \lambda \hat{w} (0) \rhoO^\kappa -\kappa, \qquad \rhoO^\kappa :=  \int_{k\in \R ^d} \frac{1}{e^{\frac{|k|^2+\kappa}{T}} -1} dk.
\end{equation}
Alors, si $\kappa$ est assez grand (ind\'ependement de $T$ et $\lambda$), il existe une fonction $V_\infty$ satisfaisant
$$\frac{V}{2}\le V_\infty -\kappa \le \frac{3V}2$$
telle que 
$$ \lim_{T\to\ii}\norm{\frac{V_T-V_{\infty}}{V}}_{L^\ii(\R^d)}=0$$
et
\begin{align} \label{eq:VT-Vinf-Sp}
\lim_{T\to\ii}\Tr \left| (-\Delta+V_T)^{-1} -(-\Delta+V_\infty)^{-1}  \right|^p =0
\end{align} 
avec $p$ le meilleur exposant tel que~\eqref{eq:p trace} ait lieu.
\end{enumerate}
\end{theorem}

Quelques commentaires:

\smallskip

\noindent(1) Ici et dans la suite on assimile un op\'erateur \`a trace $\gamma$ et son noyau int\'egral $\gamma (x;y)$. La partie diagonale $\gamma (x;x)$ peut se d\'efinir plus pr\'ecis\'ement par 
$$ \gamma (x;x) := \sum_j \lambda_j |u_j(x)| ^2$$
o\`u $\lambda_j,u_j$ sont les valeurs et fonctions propres de $\gamma$.
 
\smallskip

\noindent(2) On obtient l'\'energie libre~\eqref{eq:rHF} en ins\'erant un \'etat quasi-libre~\eqref{eq:quasi libre} dans l'\'energie libre compl\`ete~\eqref{eq:grand can free func} et en n\'egligeant le terme d'\'echange. En fait, un \'etat quasi-libre $\Gamma$ est enti\`erement d\'etermin\'e par sa premi\`ere matrice de densit\'e r\'eduite $\gamma$, et son \'energie libre est donn\'ee par 
 $$ 
 \cF_\lambda [\Gamma] = \cF^{\rm rH}[\gamma] + \frac{\lambda}{2} \iint w(x-y) |\gamma(x;y)|^2 dx dy. 
 $$
Le dernier terme est retir\'e pour simplifier l'approche de preuve du Th\'eor\`eme~\ref{thm:contre}, et parce qu'il sera pris en compte de mani\`ere diff\'erente (comparer avec la Remarque~\ref{rem:direct exchange}).

\smallskip

\noindent(3) Derri\`ere le choix de potentiel chimique~\eqref{eq:rhoOkappa} se trouve l'intuition que la densit\'e de mati\`ere des \'etats qui nous int\'eressent diverge, certes, mais de mani\`ere uniforme en espace, avec le taux $\rhoO^\kappa$. Ainsi, il est possible de compenser cette divergence en ajustant le potentiel chimique, c'est \`a dire en p\'enalisant la densit\'e de particules globale.

\smallskip

\noindent(4) La deuxi\`eme partie du th\'eor\`eme (la plus difficile) est celle qui donne le Hamiltonien \`a une particule renormalis\'e
$$ h_{\infty} = - \Delta + V_\infty$$
qui nous servira par la suite. 

\medskip

\noindent\textbf{Limite des \'etats de Gibbs quantiques.} Nous avons maintenant d\'efini tous les objets n\'ecessaires pour \'etudier la limite des \'etats de Gibbs quantiques~\eqref{eq:grand can Gibbs} en 2D. Comme d\'ej\`a mentionn\'e, nous ne pouvons traiter le cas Hilbert-Schmidt en toute g\'en\'eralit\'e. L'\'enonc\'e suivant est une version simplifi\'ee des r\'esultats principaux de~\cite{LewNamRou-18c}:

\begin{theorem}[\textbf{Limite de champ moyen \`a temp\'erature positive en 2D}]\label{thm:main HS}

\flushleft Soit $\Gamma_\lambda$ l'\'etat de Gibbs grand-canonique associ\'e \`a~\eqref{eq:N body Schro} avec $d=2$ et 
$$ h = - \Delta + V = - \Delta + |x| ^s$$
avec $s$ assez grand (ce qui implique que~\eqref{eq:p trace} a lieu avec $p$ suffisament proche de $1$, voir Exemple~\ref{exe:s,p,d}).

On suppose que $w\in L^1(\R^2)$ est une fonction paire satisfaisant 
\begin{equation}\label{eq:interaction}
\widehat w(k)\ge 0, \quad \int_{\R^2} \widehat w(k)  \left( 1 + |k|^{1/2}  \right)  dk <\infty,\qquad \int_{\R^2}|w(x)|V(x)^2\,dx<\ii.
\end{equation}
On fixe le potentiel chimique comme dans~\eqref{eq:rhoOkappa}:
\begin{equation}\label{eq:fix nu}
\nu = \lambda \hat{w} (0) \rhoO^\kappa -\kappa
\end{equation}
avec $\kappa$ une constante assez grande.

Soit $\gamma^{\rm rHF}$ la matrice de densit\'e r\'eduite d\'efinie par le Th\'eor\`eme~\ref{thm:contre}, et fixons dans~\eqref{eq:grand can free func}
$$ 
E_0 = \frac{\lambda}{2}\iint_{\R^2 \times \R^2} \gamma^{\rm rHF} (x;x) w(x-y) \gamma^{\rm rHF} (y;y) dxdy.
$$

Soient $\mu_0$ et $\mu$ les mesure de Gibbs gaussienne et non-lin\'eaire d\'efinies aux Lemmes~\ref{lem:gaussienne} et~\ref{lem:NL Gibbs HS}, avec cette fois 
$$ h = h_\infty = -\Delta + V_\infty $$
comme d\'efini par le Th\'eor\`eme~\ref{thm:contre}. 

On a alors, dans la limite $T\to \infty$, $\lambda\sim T ^{-1}$
\begin{enumerate}
 \item Asymptotique de l'\'energie libre: avec $F_\lambda$ l'\'energie libre associ\'ee \`a~\eqref{eq:grand can Gibbs}, $\tilde{F}_0$ celle associ\'ee \`a~\eqref{eq:quasi-libre ref}, et $z_r$ la fonction de partition relative classique d\'efinie au Lemme~\ref{lem:NL Gibbs HS}
 \begin{equation}\label{eq:main 2D ener libre}
\frac{F_\lambda - \tilde{F}_0}{T} \underset{T\to \infty}{\longrightarrow} - \log z_r.  
 \end{equation}
\item Asymptotique des matrices de densit\'es r\'eduites: soit $\Gamma_\lambda ^{(k)}$ la $k$-\`eme matrice de densit\'e r\'eduite de l'\'etat de Gibbs~\eqref{eq:grand can Gibbs}. On a, pour tout $k\in \N$, 
\begin{equation}\label{eq:main 2D states}
\tr_{\gH_k} \left| \frac{k!}{T^k} \Gamma_\lambda ^{(k)} - \int |u ^{\otimes k} \rangle \langle u ^{\otimes k} | d\mu(u) \right| ^2 \underset{T\to \infty}{\longrightarrow} 0.  
\end{equation}
\item Asymptotique pour la premi\`ere matrice de densit\'e r\'eduite relative: 
\begin{equation}\label{eq:main 2D states rel}
\tr_{\gH} \left| \frac{1}{T} \left(\Gamma_\lambda ^{(1)} - \gamma^{\rm rHF} \right) - \left( \int |u \rangle \langle u  | d\mu(u) - \int |u \rangle \langle u  | d\mu_0 (u) \right) \right|  \underset{T\to \infty}{\longrightarrow} 0.  
\end{equation}
\end{enumerate} 
\end{theorem}

Quelques commentaires:

\smallskip

\noindent(1) On trouvera dans l'article original~\cite{LewNamRou-18c} des bornes explicites sur $s,p$ sous lesquelles nous pouvons d\'emontrer le th\'eor\`eme. Elles sont loin d'\^etre optimales, c'est pourquoi j'ai pr\'ef\'er\'e pr\'esenter le th\'eor\`eme sous la forme simplifi\'ee ci-dessus.

\smallskip

\noindent(2) Il est ici fondamental de bien choisir l'\'etat quasi-libre de r\'ef\'erence, de lui associer un hamitonien renormalis\'e et la mesure gaussienne correspondante. Cette mesure et la mesure sans interaction sont mutuellement singuli\`eres.

\smallskip

\noindent(3) Le choix de potentiel chimique~\eqref{eq:fix nu} a deux parties: le premier terme diverge de fa\c{c}on \`a compenser la divergence de la partie directe des interactions, le second donne le potentiel chimique de la th\'eorie classique limite.
 
\smallskip

\noindent(4) L'asymptotique d'\'energie libre~\eqref{eq:main 2D ener libre} indique que la contribution de loin la plus grande est donn\'ee par le probl\`eme r\'eduit aux \'etats quasi-libres du Th\'eor\`eme~\ref{thm:contre}. Calculer l'ordre suivant, $-T \log z_r$ est toutefois fondamental pour trouver la mesure limite correcte. Essentiellement, savoir que le membre de gauche de~\eqref{eq:main 2D ener libre} est born\'e indique seulement que la mesure limite est absolument continue par rapport \`a $\mu_0$.
 
\smallskip

\noindent(5) On s'attend \`a ce que~\eqref{eq:main 2D states} ne soit pas seulement une convergence en norme Hilbert-Schmidt, mais dans la classe de Schatten associ\'ee au plus petit $p$ satisfaisant~\eqref{eq:p trace}. Ceci reste un probl\`eme ouvert, sauf si $k$ est assez petit, voir~\cite{LewNamRou-18c}.
 
\smallskip

\noindent(6) Le dernier r\'esultat~\eqref{eq:main 2D states rel} est important en pratique car la norme de trace donne la topologie naturelle sur les matrices de densit\'e r\'eduites. C'est toutefois seulement la matrice de densit\'e relative (celle du probl\`emes complet moins celle de l'\'etat quasi-libre de r\'ef\'erence) qui peut converger en norme de trace.

\smallskip

\noindent(7) L'\'enonc\'e du pendant de ce th\'eor\`eme pour le cas o\`u les particules sont confin\'ees dans une bo\^ite p\'eriodique est beaucoup plus simple, car l'\'etat quasi-libre de r\'ef\'erence est dans ce cas \'egal \`a l'\'etat de Gibbs sans interactions. Voir~\cite{LewNamRou-18b} pour une pr\'esentation r\'esum\'ee de ce cas.

\bigskip

\noindent \textbf{Remerciements.} Ce projet de recherche a \'et\'e financ\'e par le programme de Recherche et d'Innovation ``Horizon 2020'' du Conseil Europ\'een de la Recherche (ERC) (Bourses de Recherche MDFT No 725528 et CORFRONMAT No 758620), par le ``People Programme / Marie Curie Actions'' (European
Community's Seventh Framework Programme, REA Grant Agreement 291734) et par l'ANR
(Projets NoNAP ANR-10-BLAN-0101 \& Mathostaq ANR-13-JS01-0005-01). Nous avons b\'en\'efici\'e de discussions inspirantes avec J\"urg Fr\"ohlich, Markus Holzmann, Antti Knowles, Benjamin Schlein, Vedran Sohinger, Robert Seiringer, Laurent Thomann et Jakob Yngvason. Grand merci \'egalement aux organisateurs du congr\`es de la Soci\'et\'e Math\'ematique de France (Lille, Juin 2018).

\renewcommand{\refname}{R\'ef\'erences}

\begin{thebibliography}{10}

\bibitem{ArnMoo-01}
{\sc P.~Arnold and G.~Moore}, {\em {BEC} transition temperature of a dilute
  homogeneous imperfect {B}ose gas}, Phyical Review Letters, 87 (2001),
  p.~120401.

\bibitem{BayBlaiHolLalVau-99}
{\sc G.~Baym, J.-P. Blaizot, M.~Holzmann, F.~Lalo{\"e}, and D.~Vautherin}, {\em
  The transition temperature of the dilute interacting {B}ose gas}, Physical
  Review Letters, 83 (1999), pp.~1703--1706.

\bibitem{BayBlaiHolLalVau-01}
\leavevmode\vrule height 2pt depth -1.6pt width 23pt, {\em {Bose-Einstein
  transition in a dilute interacting gas}}, European Physical Journal B, 24
  (2001), p.~107.

\bibitem{BenPorSch-15}
{\sc N.~{Benedikter}, M.~{Porta}, and B.~{Schlein}}, {\em {Effective Evolution
  Equations from Quantum Dynamics}}, Springer Briefs in Mathematical Physics,
  Springer, 2016.

\bibitem{BetUel-10}
{\sc V.~Betz and D.~Ueltschi}, {\em Critical temperature of dilute {B}ose
  gases}, Phys. Rev. A, 81 (2010), p.~023611.

\bibitem{BisDavSimBla-09}
{\sc R.~N. Bisset, M.~J. Davis, T.~P. Simula, and P.~B. Blakie}, {\em
  Quasicondensation and coherence in the quasi-two-dimensional trapped {B}ose
  gas}, Phys. Rev. A, 79 (2009), p.~033626.

\bibitem{Bourgain-96}
{\sc J.~Bourgain}, {\em Invariant measures for the {2D}-defocusing nonlinear
  {S}chr\"odinger equation}, Comm. Math. Phys., 176 (1996), pp.~421--445.

\bibitem{Bourgain-97}
\leavevmode\vrule height 2pt depth -1.6pt width 23pt, {\em {Invariant measures
  for the Gross-Pitaevskii equation}}, Journal de Math\'ematiques Pures et
  Appliqu\'ees, 76 (1997), pp.~649--02.

\bibitem{BurThoTzv-09}
{\sc N.~{Burq}, L.~{Thomann}, and N.~{Tzvetkov}}, {\em {Gibbs measures for the
  non linear harmonic oscillator}}, in Journ\'ees EDP \'Evian 2009, 2009.

\bibitem{BurThoTzv-10}
\leavevmode\vrule height 2pt depth -1.6pt width 23pt, {\em {Long time dynamics
  for the one dimensional non linear Schr\"odinger equation}}, Ann. Inst.
  Fourier., 63 (2013), pp.~2137--2198.

\bibitem{CacSuz-14}
{\sc F.~Cacciafesta and A.-S. {de Suzzoni}}, {\em Invariant measure for the
  {S}chr\"odinger equation on the real line}, J. Func Anal, 269 (2015),
  pp.~271--324.

\bibitem{PraDeb-03}
{\sc G.~da~Prato and A.~Debussche}, {\em Strong solutions to the stochastic
  quantization equations}, Ann. Probab., 32 (2003), pp.~1900--1916.

\bibitem{BouDebFuk-17}
{\sc A.~de~Bouard, A.~Debussche, and R.~Fukuizumi}, {\em Long time behavior of
  {G}ross-{P}itaevskii equation at positive temperature}.
\newblock arXiv:1708.01961, 2017.

\bibitem{DerGer-13}
{\sc J.~Derezi{\'n}ski and C.~G{\'e}rard}, {\em {Mathematics of Quantization
  and Quantum Fields}}, Cambridge University Press, Cambridge, 2013.

\bibitem{DeuSeiYng-18}
{\sc A.~Deuchert, R.~Seiringer, and J.~Yngvason}, {\em Bose-{E}instein
  condensation in a dilute, trapped gas at positive temperature}.
\newblock arXiv:1803.05180.

\bibitem{FroKnoSchSoh-16}
{\sc J.~Fr\"ohlich, A.~Knowles, B.~Schlein, and V.~Sohinger}, {\em Gibbs
  measures of nonlinear {S}chr\"odinger equations as limits of quantum
  many-body states in dimensions $d\leq 3$}, Communications in Mathematical
  Physics, 356 (2017), pp.~883--980.

\bibitem{FroKnoSchSoh-17}
\leavevmode\vrule height 2pt depth -1.6pt width 23pt, {\em A microscopic
  derivation of time-dependent correlation functions of the {1D} cubic
  nonlinear {Schr\"odinger} equation}.
\newblock arXiv:1703.04465, 2017.

\bibitem{GioCarCas-07}
{\sc L.~Giorgetti, I.~Carusotto, and Y.~Castin}, {\em Semiclassical field
  method for the equilibrium {B}ose gas and application to thermal vortices in
  two dimensions}, Phys. Rev. A, 76 (2007), p.~013613.

\bibitem{GliJaf-87}
{\sc J.~Glimm and A.~Jaffe}, {\em Quantum Physics: A Functional Integral Point
  of View}, Springer-Verlag, 1987.

\bibitem{Golse-13}
{\sc F.~{Golse}}, {\em {On the Dynamics of Large Particle Systems in the Mean
  Field Limit}}, ArXiv e-prints 1301.5494,  (2013).
\newblock Lecture notes for a course at the NDNS+ Applied Dynamical Systems
  Summer School "Macroscopic and large scale phenomena", Universiteit Twente,
  Enschede (The Netherlands).

\bibitem{Gottlieb-05}
{\sc A.~D. Gottlieb}, {\em Examples of bosonic de {F}inetti states over finite
  dimensional {H}ilbert spaces}, J. Stat. Phys., 121 (2005), pp.~497--509.

\bibitem{Hairer-14}
{\sc M.~Hairer}, {\em A theory of regularity structures}, Inventiones
  Mathematicae, 198 (2014), pp.~269--504.

\bibitem{HolBay-03}
{\sc M.~Holzmann and G.~Baym}, {\em Condensate density and superfluid mass
  density of a dilute {Bose-Einstein} condensate near the condensation
  transition}, Physical Review Letters, 90 (2003), p.~040402.

\bibitem{HolCheKra-08}
{\sc M.~Holzmann, M.~Chevallier, and W.~Krauth}, {\em Universal correlations
  and coherence in quasi-two-dimensional trapped {B}ose gases}, Europhys.
  Lett., 82 (2008), p.~30001.

\bibitem{HolKra-08}
{\sc M.~Holzmann and W.~Krauth}, {\em {Kosterlitz-Thouless} transition of the
  quasi-two-dimensional trapped {B}ose gas}, Phys. Rev. Lett., 100 (2008),
  p.~190402.

\bibitem{KasProSvi-01}
{\sc V.~A. Kashurnikov, N.~V. Prokof'ev, and B.~V. Svistunov}, {\em Critical
  temperature shift in weakly interacting {B}ose gas}, Physical Review Letters,
  87 (2001), p.~120402.

\bibitem{Knowles-thesis}
{\sc A.~Knowles}, {\em Limiting dynamics in large quantum systems.}
\newblock Doctoral thesis, ETH Z\"urich, 2009.

\bibitem{Kupiainen-16}
{\sc A.~Kupiainen}, {\em Renormalization group and stochastic pdes}, Annales
  Henri Poincar\'e, 17 (2016), pp.~497--535.

\bibitem{LebRosSpe-88}
{\sc J.~L. Lebowitz, H.~A. Rose, and E.~R. Speer}, {\em Statistical mechanics
  of the nonlinear {S}chr\"odinger equation}, J. Statist. Phys., 50 (1988),
  pp.~657--687.

\bibitem{LeiMyr-77}
{\sc J.~M. {Leinaas} and J.~{Myrheim}}, {\em {On the theory of identical
  particles}}, Nuovo Cimento B Serie, 37 (1977), pp.~1--23.

\bibitem{Lewin-ICMP}
{\sc M.~Lewin}, {\em {Mean-Field limit of Bose systems: rigorous results}},
  Preprint (2015) arXiv:1510.04407.

\bibitem{Lewin-XEDP-12}
\leavevmode\vrule height 2pt depth -1.6pt width 23pt, {\em Gaz de bosons dans
  le r{\'e}gime de champ moyen~: les th{\'e}ories de {H}artree et
  {B}ogoliubov}, in S{\'e}minaire {L}aurent {S}chwartz -- {EDP} et
  applications, IH\'ES, 2012-2013.
\newblock Exp. no 3.

\bibitem{LewNamRou-14d}
{\sc M.~Lewin, P.~Nam, and N.~Rougerie}, {\em Derivation of nonlinear {G}ibbs
  measures from many-body quantum mechanics}, Journal de l'Ecole Polytechnique,
  2 (2016), pp.~553--606.

\bibitem{LewNamRou-18c}
\leavevmode\vrule height 2pt depth -1.6pt width 23pt, {\em {Classical field
  theory limit of 2D many-body quantum Gibbs states}}.
\newblock arXiv:1810.08370, 2018.

\bibitem{LewNamRou-17}
\leavevmode\vrule height 2pt depth -1.6pt width 23pt, {\em Gibbs measures based
  on {1D} (an)harmonic oscillators as mean-field limits}, Journal of
  Mathematical Physics, 59 (2018).

\bibitem{LewNamRou-18b}
\leavevmode\vrule height 2pt depth -1.6pt width 23pt, {\em {The interacting 2D
  Bose gas and nonlinear Gibbs measures}}.
\newblock arXiv:1805.03506, 2018.
\newblock Oberwolfach Abstract.

\bibitem{LieSei-09}
{\sc E.~H. Lieb and R.~Seiringer}, {\em The {S}tability of {M}atter in
  {Q}uantum {M}echanics}, Cambridge Univ. Press, 2010.

\bibitem{LieSeiSolYng-05}
{\sc E.~H. Lieb, R.~Seiringer, J.~P. Solovej, and J.~Yngvason}, {\em The
  mathematics of the {B}ose gas and its condensation}, Oberwolfach {S}eminars,
  Birkh{\"a}user, 2005.

\bibitem{MouWeb-15}
{\sc J.-C. Mourrat and H.~Weber}, {\em Global well-posedness of the dynamic
  {$\Phi^4$} model in the plane}, Annals of Probability, 45 (2015).

\bibitem{Nelson-66}
{\sc E.~Nelson}, {\em A quartic interaction in two dimensions}, in Mathematical
  Theory of Elementary Particles, MIT Press, Cambdridge, Mass., 1966,
  pp.~69--73.

\bibitem{OhTho-15}
{\sc T.~Oh and L.~Thomann}, {\em {A pedestrian approach to the invariant Gibbs
  measures for the 2D defocusing nonlinear Schr\"odinger equations}}.
\newblock arXiv:1509.02093, 2015.

\bibitem{ProRueSvi-01}
{\sc N.~Prokof'ev, O.~Ruebenacker, and B.~Svistunov}, {\em Critical point of a
  weakly interacting two-dimensional {B}ose gas}, Phys. Rev. Lett., 87 (2001),
  p.~270402.

\bibitem{ProSvi-02}
{\sc N.~V. Prokof'ev and B.~V. Svistunov}, {\em Two-dimensional weakly
  interacting {B}ose gas in the fluctuation region}, Physical Review A, 66
  (2002), p.~043608.

\bibitem{RocZhuZhu-16}
{\sc M.~R\"ockner, R.~Zhu, and X.~Zhu}, {\em Ergodicity for the stochastic
  quantization problems on the {2D}-torus}, Communications in Mathematical
  Physics, 352 (2017), pp.~1061--1090.

\bibitem{Rougerie-LMU}
{\sc N.~Rougerie}, {\em {De Finetti theorems, mean-field limits and
  Bose-Einstein condensation}}.
\newblock arXiv:1506.05263, 2014.
\newblock LMU lecture notes.

\bibitem{Rougerie-spartacus}
\leavevmode\vrule height 2pt depth -1.6pt width 23pt, {\em Th{\'e}or{\`e}mes de
  De Finetti, limites de champ moyen et condensation de Bose-Einstein}, Les
  cours Peccot, Spartacus IDH, Paris, 2016.
\newblock Cours Peccot, Coll{\`e}ge de France : f{\'e}vrier-mars 2014.

\bibitem{Schlein-08}
{\sc B.~Schlein}, {\em Derivation of effective evolution equations from
  microscopic quantum dynamics}, arXiv eprints,  (2008).
\newblock Lecture Notes for a course at ETH Zurich.

\bibitem{Seiringer-06}
{\sc R.~Seiringer}, {\em A correlation estimate for quantum many-body systems
  at positive temperature}, Rev. Math. Phys., 18 (2006), pp.~233--253.

\bibitem{Seiringer-08}
\leavevmode\vrule height 2pt depth -1.6pt width 23pt, {\em Free energy of a
  dilute {B}ose gas: Lower bound}, Comm. Math. Phys., 279 (2008), pp.~596--636.

\bibitem{SeiUel-09}
{\sc R.~Seiringer and D.~Ueltschi}, {\em Rigorous upper bound on the critical
  temperature of dilute {B}ose gases}, Phys. Rev. B, 80 (2009), p.~014502.

\bibitem{Simon-74}
{\sc B.~Simon}, {\em The {$P(\Phi )_{2}$} {E}uclidean (quantum) field theory},
  Princeton University Press, Princeton, N.J., 1974.
\newblock Princeton Series in Physics.

\bibitem{Simon-Nelson}
\leavevmode\vrule height 2pt depth -1.6pt width 23pt, {\em {Ed Nelson's} work
  in quantum theory}, in Diffusion, Quantum Theory, and Radically Elementary
  Mathematics, Princeton University Press, 2006, pp.~75--93.

\bibitem{SimDavBlak-08}
{\sc T.~P. Simula, M.~J. Davis, and P.~B. Blakie}, {\em Superfluidity of an
  interacting trapped quasi-two-dimensional {B}ose gas}, Phys. Rev. A, 77
  (2008), p.~023618.

\bibitem{Summers-12}
{\sc S.~J. {Summers}}, {\em {A Perspective on Constructive Quantum Field
  Theory}}.
\newblock arXiv:1203.3991, 2012.

\bibitem{TsaWeb-16}
{\sc P.~Tsatsoulis and H.~Weber}, {\em Spectral gap for the stochastic
  quantization equation on the 2-dimensional torus}, Annales de l'Institut
  Henri Poincar\'e,  (2018).

\bibitem{Yin-10}
{\sc J.~Yin}, {\em Free energies of dilute {B}ose gases: Upper bound}, Journal
  of Statistical Physics, 141 (2010), p.~683.

\end{thebibliography}

\end{document}